\documentstyle{article}
\begin{document}
\newcommand{\lsim}{\:\raise -4pt\hbox{$\stackrel{\textstyle <} {\sim}$}\:}
\begin{center}
\Large\bf{Production of mini-(gluon)jets and strangeness 
enhancement in pA and AA collisions at relativistic energies}
 \\
\vspace{1cm}\large
  Tai An$^2$ and Sa Ben-Hao$^{1,3,4}$ \\
\begin{tabbing}
ttttt \= tt \= \kill
\>1. CCAST (World Lab.), P. O. Box 8730 Beijing, 100080 \\
\> \>China. \\
\>2. Institute of High Energy Physics, Academia Sinica, \\
\> \> P. O. Box 918, Beijing, 100039 China.\footnotemark \\       
\>3. China Institute of Atomic Energy, P. O. Box 275 (18), \\
\> \>Beijing, 102413 China.\\
\>4. Institute of Theoretical Physics, Academia Sinica, \\
\> \>Beijing 100080 China.\\

\end{tabbing}
\end{center}
\footnotetext{mailing address. \\Email: taian@hptc1.ihep.ac.cn}
\normalsize
\large
\begin{abstract}
The idea that effective string tension increases as a result of the hard gluon 
kinks on a string is applied to study the strange particle production in 
proton-nucleus and nucleus-nucleus collisions. It is found that the effective 
string tension increases with the increase of centrality and mass of 
the colliding system as a consequence of the mini-(gluon)jet production 
stemming from the collective string-string interaction. This mechanism leads to 
strangeness enhancement in pA and AA collisions through the enhanced 
production of the strange quark pairs from the color field of strings. We 
discuss different roles played by this mechanism and rescattering of the final 
state hadrons in the production of strange particles and compare our results 
with experimental data.


\end{abstract}
\baselineskip 0.6cm
\parindent=0.3cm
\parskip=0.3cm
\hspace{0.3cm}
\newpage
\section{Introduction}
Strangeness as a possible signature of the phase transition from a hadronic 
state to a QGP state was put forward about 15 years ago \cite{rafelski1}. It
was based on the prediction that the production of strange quark pairs would 
be enhanced as a result of the approximate chiral symmetry restoration in a 
QGP state in comparison with a hadronic state. The strangeness enhancement in 
pA and AA collisions with respect to the nucleon-nucleon collision has been 
investigated and confirmed by many experimental groups \cite{wa85}\cite{na351}
\cite{na36}\cite{naga}. However, alternative explanations exist in the 
hadronic regime, like rescattering, string-string interaction, etc.  
\cite{satai}\cite{venus}\cite{rqmd}. The first detailed theoretical study of 
strangeness production can be found in \cite{rafelski2}.

We have done a series of studies in recent years investigating strangeness 
enhancement based on a rescattering scheme \cite{satai}\cite{satai2}
\cite{satai3}, from which a Monte-Carlo event generator, LUCIAE, was developed 
\cite{luciae}. Those studies indicate that including rescattering of the final 
state hadrons is still not enough to reproduce the NA35 data of strange 
particle production, which imply enhanced production of strange quark pairs in 
nucleus-nucleus collisions. To reproduce the NA35 data needs further to rely 
on the reduction of strange quark suppression in nucleus-nucleus collisions 
comparing to the nucleon-nucleon collision \cite{satai2}\cite{satai3}. 
Similarly, in order to reproduce the NA35 data, the RQMD generator, equipped 
with rescattering though, has to resort to the colour rope mechanism \cite
{rqmd}. In this picture it is assumed that the neighboring interacting strings 
might form a string cluster called colour rope in pA and AA collisions. The 
colour rope then fragments in a collective way and tends to enhance the 
production of the strange quark pairs from the colour field of strings through 
the increase of the effective string tension. 
  
It has been known for years that the strange quark suppression factor 
($\lambda$ hereafter), i.e. the suppression of s quark pair production in the 
color field with respect to u or d pair production, in hadron-hadron 
collisions is not a constant, but energy-dependent, increasing from a value of 
0.2 at the ISR energies to about 0.3 at the top of the SPS energies \cite{kapa}.
In \cite{kapat} we proposed a scenario to investigate the energy dependence of 
$\lambda$ in hh collisions by relating the effective string tension to the 
production of hard gluon jets (mini-jets). A parametrization form was then 
obtained, which reproduces the energy dependence of $\lambda$ in hh collisions 
reasonably well. By taking the energy dependence of $\lambda$ (and the other 
parameters related to the effective string tension) into account our model 
reproduce nicely the data of strange particle production in hh collisions.

We have in \cite{satai2}\cite{satai3} described successfully the NA35 
data of pA and AA collisions based on the idea of reduction of strangeness 
suppression via  adjusting the concerned parameters $\lambda$ etc. 
relating to the effective string tension. The relations between the reduction 
of strangeness suppression (the parameter $\lambda$ etc.) and effective 
string tension are thus urgent to be established. In this work we use the idea of 
\cite{kapat} and the Firecracker model \cite{fire} to study above relations. 
The study reveals that the mini-jet production from the string-string 
interaction might play a role in the strangeness production. It causes the 
strange quark suppression factor increasing with centrality and mass of the 
colliding system in nucleus-nucleus collisions in addition to increasing with 
energy (such kind of the strange quark suppression factor increasing with 
energy, centrality, and mass of the colliding system is called as reduction 
mechanism of s quark suppression hereafter). This study provides then a 
dynamic explanation for the reduction of strangeness suppression shown in the 
experimental data of nucleus-nucleus collisions.
 
\section{Brief review of LUCIAE model}
LUCIAE model is developed based on the FRITIOF model \cite{fritiof}. FRITIOF
is a string model, which started from the modeling of inelastic hadron-hadron
collisions and it has been successful in describing many experimental data
from the low energies at the ISR-regime all the way to the SPS energies 
\cite{B.N} \cite{H.P1}. In this model a hadron is assumed to behave 
like a massless relativistic string corresponding to a confined color 
force field of a vortex line character embedded in a type II color 
superconducting vacuum. A hadron-hadron collision is pictured as the multi-
scatterings of the partons inside the two colliding hadrons. In FRITIOF, 
during the collision two hadrons are excited due to longitudinal momentum 
transfers and/or a Rutherford Parton Scattering (RPS). The highly excited 
states will emit bremsstrahlung gluons according to the soft radiation model. 
They are afterwards treated as excitations i.e. the Lund Strings and  
allowed to decay into final state hadrons according to the Lund
fragmentation scheme \cite{lund}.

The FRITIOF model has been extended to also describe hadron-nucleus and 
nucleus-nucleus collisions by assuming that the reactions are superpositions 
of binary hadron-hadron collisions in which the geometry of the nucleus
plays an important role because the nuclei should then behave as a
``frozen'' bag of nucleons. 

However in the relativistic nucleus-nucleus collision there are generally 
many excited strings formed close by each other during a collision. Thus in 
LUCIAE a Firecracker model is proposed to deal with the string-string 
collective interaction. In the Firecracker model it is assumed that several strings 
from a relativistic heavy ion reaction will form a cluster and then the 
strings inside such a cluster will interact in a collective way.  We assume
that the groups of neighbouring strings in a cluster may form interacting 
quantum states so that both the emission of gluonic bremsstrahlung as well as 
the fragmentation properties can be affected by the large common energy 
density, see \cite{fire} for the details. 

In relativistic nucleus-nucleus collision there are generally a lot of 
hadrons produced, however FRITIOF does not include the final state 
interactions. Therefore in LUCIAE a rescattering model is devised to consider   
the interactions of produced hadrons with each other and with the surrounding 
cold spectator matter. The distributions of the final state hadrons will be 
affected by the rescattering process. The details have been described in 
\cite{satai2} and \cite{luciae}, here we just give the complete list of the 
reactions including in LUCIAE, which are cataloged into:
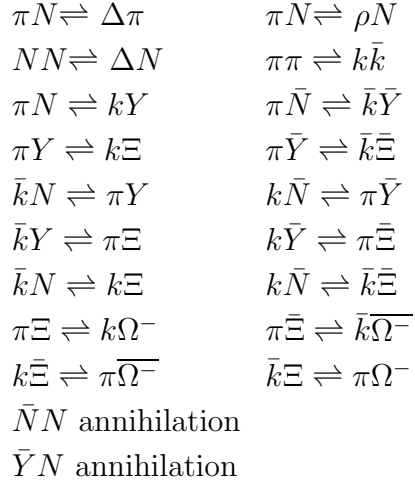
\begin{tabbing}
ttttttttttttttt\=ttttttttttttttt\=tttttt\=tttttttttttttttt\=  \kill
\>$\pi$$N$$\rightleftharpoons$ $\Delta$$\pi$
\> \>$\pi$$N$$\rightleftharpoons$ $\rho$$N$\\
\> $N$$N$$\rightleftharpoons$ $\Delta$$N$
\> \>$\pi\pi \rightleftharpoons k\bar{k}$\\
\>$\pi N \rightleftharpoons kY$
\> \>$\pi\bar{N} \rightleftharpoons  \bar{k}\bar{Y}$\\
\>$\pi Y  \rightleftharpoons k\Xi$
\> \>$\pi\bar{Y}  \rightleftharpoons  \bar{k}\bar{\Xi}$\\
\>$\bar{k}N  \rightleftharpoons  \pi Y$ 
\> \>$k\bar{N}  \rightleftharpoons  \pi\bar{Y}$\\
\>$\bar{k}Y  \rightleftharpoons  \pi\Xi$
\> \>$k\bar{Y}  \rightleftharpoons  \pi\bar{\Xi}$\\
\>$\bar{k}N  \rightleftharpoons  k\Xi$
\> \>$k\bar{N}  \rightleftharpoons  \bar{k}\bar{\Xi}$\\
\>$\pi\Xi \rightleftharpoons k\Omega^- $
\> \>$\pi\bar{\Xi} \rightleftharpoons  \bar{k}\overline{\Omega^-}$\\
\>$k\bar{\Xi} \rightleftharpoons \pi\overline{\Omega^-}$
\> \>$\bar{k}\Xi \rightleftharpoons \pi\Omega^-$\\
\>$\bar{N}N$ annihilation\\
\>$\bar{Y}N$ annihilation\\
\end{tabbing}
where $Y$ refers to the $\Lambda$ or $\Sigma$ and $\Xi$ refers to the $\Xi^-$ 
or $\Xi^0$. There are 364 reactions involved altogether.  

\section{Results and discussions}
String tension is defined as the energy per unit length of the string. 
However, the existence of gluons on the string (regarded as the transverse 
excitation or `kink' on a string in the Lund string fragmentation model) would 
wrinkle a string and give a fractal structure. Such a wrinkled string has 
larger energy density in comparison with a string without gluon, thereby an 
enhanced string tension effectively \cite{torbjon}.
  
The following form has been used in \cite{kapat} to parametrize the relation
between the effective string tension and the hard gluon jets on a string 
\begin{equation}
\kappa_{eff}=\kappa_{0} (1-\xi)^{-\alpha},
\label{f1}
\end{equation} 
where $\kappa_{0}$ is the string tension of the pure $q\bar{q}$ string, 
$\alpha$ is a parameter to be determined by experiments and $\xi$ is 
calculated by
\begin{equation}
\xi =\frac{\ln(\frac{k_{\perp max}^2}{s_{0}})}{\ln (\frac{s}{s_{0}}) + 
\sum_{j=2}^{n-1} \ln (\frac{k_{\perp j}^2}{s_{0}})},
\label{f2}
\end{equation}
which represents the scale that a multigluon string is deviated from a pure 
$q\bar{q}$ string. Here the multigluon string state has (n-2) gluons, 
indexed in a colour connected way from the $q$ (index 1) to the $\bar{q}$ 
(index n) and $k_{\perp j}$, j=2,...,(n-1), are the transverse momenta of the 
emitted gluons with $k_{\perp j}^2 \geq s_{0}$. The parameter $\sqrt{s_{0}}$ 
is of the order of a typical hadron mass. The parameter $\alpha$ in Eq.(\ref
{f1}) and the $\sqrt{s_{0}}$ in Eq.(\ref{f2}) are determined by hh data to be 
about 3.5 and 0.8 GeV, respectively \cite{kapat}.

In the Lund string fragmentation model, the $q\bar{q}$ pairs with the quark 
mass $m$ and the transverse momentum $p_{t}$ are produced from the colour 
field of a string by a quantum tunneling process with probability
\begin{equation}
\exp(\frac{-\pi m^{2}}{\kappa_{eff}})\exp(\frac{-\pi p_{t}^{2}}{\kappa_{eff}}).
\label{f3}
\end{equation}
The above equation shows that the probability of the $s\bar{s}$ pair 
production with respect to a $u\bar{u}$ (or $d\bar{d}$) pair as well as the 
probability of a high $p_{t}$ $q\bar{q}$ pair production will be enhanced in a 
field with larger $\kappa_{eff}$. 

Assume that the width of the Gaussian transverse momentum distribution of 
$q\bar{q}$ pairs and the strangeness suppression factor of a string with 
effective string tension $\kappa_{eff1}$ are $\sigma_{1}$ and $\lambda_{1}$, 
respectively, then those quantities of a string with effective string tension 
$\kappa_{eff2}$ can be calculated from Eq.(\ref{f3}), i.e.  
\begin{eqnarray}
\sigma_{2} &=& \sigma_{1}(\frac{\kappa_{eff2}}{\kappa_{eff1}})^{1/2}\nonumber\\
\lambda_{2} &=& \lambda_{1}^ {\frac{\kappa_{eff1}}{\kappa_{eff2}}}.
\label{f4}
\end{eqnarray}

We see that $\sigma$ and $\lambda$ for two string states are related by the 
ratio of the effective string tensions of this two string states only. 
It should be noted that the discussion above is also valid for the production 
of the diquark pairs from the string field, i.e. the production of the diquark 
pairs with respect to the $q\bar{q}$ pairs will be enhanced from a string
with larger $\kappa_{eff}$, therefore, more baryons (or antibaryons) will be 
formed in the final state. 

In JETSET routine which runs together with LUCIAE event generator, there are 
model parameters PARJ(2) (the same as $\lambda$) and PARJ(3). PARJ(3) is the 
extra suppression of strange diquark production compared to the normal 
suppression of strange quark pair. Both PARJ(2) and PARJ(3) are responsible 
for the s quark (diquark) suppression and related to the effective string 
tension. Besides $\lambda$ and PARJ(3) there is PARJ(1), which stands for the 
suppression of diquark-antidiquark pair production in the color field in 
comparison with the quark-antiquark pair production and is related to the 
effective string tension as well. How these three parameters affect the 
multiplicity distribution of final state particles can be found in 
\cite{satai2} \cite{satai3}. Another parameter PARJ(21) (the same as $\sigma$)
, which is the width of the Gaussian transverse momentum distribution of 
$q\bar{q}$ pairs in the string fragmentation, varies with $\kappa_{eff}$ too, 
but it is not related to the strangeness production directly.

It has been shown in \cite{satai2} \cite{satai3} \cite{topor} and \cite{heinz} 
that the string fragmentation by JETSET with default values of PARJ(1)=0.1, 
PARJ(2)=0.3 and PARJ(3)=0.4 overestimates the yield of strange particles in 
the pp collision at 200 GeV/c. Thus in \cite{kapat} we first retune these 
parameters by comparing with the pp data of strange particle production 
\cite{hansen}. A new set of parameters PARJ(1)=0.046, PARJ(2)=0.2, PARJ(3)=0.3 
and PARJ(21)=0.32 GeV/c (the corresponding default value is 0.37 GeV/c) are 
found for pp at 200 GeV/c. This set of parameters are then used to calculate 
the particle production in pA and AA collisions at 200 GeV/c per nucleon using 
LUCIAE event generator including the reduction mechanism of s quark 
suppression. 
 
In hh collisions there are two strings formed before fragmentation. The 
$\lambda$ values calculated above are the mean value of the two strings. When 
we talk about $\lambda$ and the other parameters in pA and AA collisions it 
also mean the corresponding values averaged over all the string states formed 
after a collision.

When LUCIAE is used to calculate the minimum bias p-nucleus collisions and the 
central sulphur-nucleus collisions at 200A GeV/c, it is found that 
$\lambda$ increases steadily with the mass of the colliding system  
from 0.22 for p+S to 0.28 for S+S and then the increase approaches gradually 
saturation see Tab.1. The values of the other three parameters in JETSET are 
also listed in Tab.1. This result is encouraging since we found in 
\cite{satai2} that in order to reproduce the NA35 data of strange production 
at 200A GeV/c, $ \lambda \simeq 0.2$ is needed for pp and p-nucleus collisions 
and $ \lambda \simeq 0.3$ for nucleus-nucleus collisions. We have repeated 
those calculations using present version of LUCIAE, the results are all close to 
the corresponding results in \cite{satai2}. In addition, we do here compare the 
NA35 data of the transverse momentum and rapidity distributions of negative 
hadrons and participant protons in central and peripheral S+S and minimum 
bias isoscalar NN collisions \cite{prl} with the corresponding results of 
LUCIAE as shown in Fig. 1, 2 and Tab. 2. The agreement between the NA35 data 
and results of LUCIAE is reasonably good except that the rapidity distributions 
of the participant protons at the target fragmentation region is lower than
the data, which might be attributed to the fact that the fragmentation 
of the target spectators is not included in our calculation.

Fig.3 shows how the $\bar{\Lambda}$ multiplicities in the full phase space 
normalized by the mean multiplicities of negatively charged particles 
calculated from LUCIAE for S+Pb at 200A GeV/c and Pb+Pb at 158A GeV/c vary
with the increase of the centrality characterized by the mean multiplicities of 
negatively charged particles. The NA35 data \cite{na351} for central S+S 
and S+Ag collisions at 200A GeV/c are also shown in this figure together with 
the corresponding results of LUCIAE. It can be seen from Fig.3 that the 
$\bar{\Lambda}$ production with respect to the negative multiplicity from the 
LUCIAE calculation increases slightly with the increasing centrality. The 
three centralities for Pb+Pb collisions are $0.0 fm <b<3.3 fm$, $3.3 fm <b<6.8
fm$ and $6.8 fm <b<10.0 fm$, respectively, the same as ones used in the NA50
experiment of Pb+Pb collisions\cite{na50} which claimed anomalous J/$\psi$ 
suppression and a hint of the QGP formation. The $\lambda$ values calculated 
by LUCIAE at three above centralities are 0.277, 0.287 and 0.290, respectively. 
Significant deviation from the LUCIAE results in Fig.3 would be expected 
if the QGP phase transition does really occur in Pb+Pb collisions since the 
approximate restoration of the chiral symmetry will mean larger $\lambda$ 
values in the QGP state.  

It is needed to point out that the increase of the effective string tension 
(hence $\lambda$) with the increase of energy in hh collisions is due to the 
production of high $k_{\perp}$-gluons from either the RPS or 
the bremsstrahlung radiation of the colour dipoles (both of which have been 
included in our calculations by using FRITIOF that runs together with LUCIAE). 
But the increase of $\lambda$ with the increase of centrality and mass of 
the colliding system at a given energy in pA and AA collisions is due to the 
collective gluon emission from the Firecracker model $-$ the more violent a 
collision, the harder the emitted Firecracker gluons, thereby the larger 
effective string tension.

Two major sources which contribute to the production of strange particles in 
LUCIAE are the $s\bar{s}$ production from the string fragmentation and the  
rescattering of the final state hadrons. We find that the rescattering 
contribution to the $\bar{\Lambda}$ production is greatly restricted due to 
the reverse reactions and the $\bar{\Lambda}N$ annihilations. For instance, 
$\bar{\Lambda}$ multiplicity in the central S+S collision at 200 A GeV/c 
will only increase by about 5\% if the cross section of the strangeness 
production reactions is doubled from its default value. By contrast, the 
strangeness production is very sensitive to $\lambda$. When a value of 
$\lambda$=0.6 is used to calculate the $\bar{\Lambda}$ production with respect 
to the negative multiplicity in the central Pb+Pb collision at 158 A GeV/c we 
find that the ratio is about 0.058 in comparison with 0.017 at $\lambda$=0.290
, enhanced by a factor of 3.4.

The success of LUCIAE model in describing the NA35 data means that present
experiments of strangeness enhancement in nucleus-nucleus collisions can be 
understood in the hadronic regime via the consideration of the collective effects 
and the final state interaction. Both the mini-jet production from the 
Firecracker model and the rescattering process affect the distributions of 
strange particles seen in experiments. 

The parameters used in LUCIAE model have been, to a large extent, fixed either 
by experimental data or by the requirement of consistency with $e^{+}e^{-}$ 
physics and hh collisions. It is encouraging that this model can 
successfully describe many experimental data in such a consistent way. We may 
be able to see the real signals of a QGP through deviations from our model 
if the data can not be accounted for within a reasonable margin of flexibility 
of the model.

In summary, we have proposed a scenario which relates the increase of the 
effective string tension to the mini-(gluon)jet production stemming from the 
string-string interaction in high energy pA and AA collisions to investigate 
the reduction of the strangeness suppression in nucleus-nucleus collisions as 
revealed by the NA35 data of strange production.  This scenario provides an 
explanation of the enhanced production of strange quark pairs from the string 
fragmentation needed to understand the NA35 data in pA and AA collisions at 
200A GeV/c.

\begin{center}Acknowledgment\end{center}
This work is partly supported by the national Natural Science Foundation of 
China.

\newpage 
\begin{center}Figure Captions\end{center}
\begin{quotation}
Fig. 1 The rapidity distributions for (a) h$^-$ and (b) participant proton.  
The N+N data and corresponding results of LUCIAE have been multiplied by 10 for 
ease of comparison. Labels are experimental data, histograms are corresponding 
results of LUCIAE. 

Fig. 2 The transverse momentum distributions for (a) h$^-$ and (b) participant 
proton. Labels are experimental data, histograms are corresponding results of 
LUCIAE.

Fig. 3 The negative multiplicity dependence of $\bar{\Lambda}$ particle in S+
Pb reaction at 200A GeV/c and Pb + Pb reaction at 158A GeV/c calculated in the 
full phase space from LUCIAE. The LUCIAE predictions for the central S+S and 
S+Ag collision at 200A GeV/c are also given in the figure. The data points are 
taken from \cite{na351}.
\end{quotation}
\newpage
\begin{tabular}{clcccc}
\multicolumn{6}{c}{Table 1. The values of four JETSET parameters in}\\
\multicolumn{6}{c}{min. bias p+A and central S+nucleus collisions at 200A GeV/c
}\\
\hline
 &p+S &p+Ag&S+S&S+Ag &S+Pb \\
 &min.bias &min.bias&central&central &central\\
\hline
PARJ(1) &0.0654&0.0725&0.107&0.109 &0.109 \\
PARJ(2)  &0.220&0.231 &0.282 &0.284 &0.284 \\
PARJ(3)  &0.323&0.328 &0.380 &0.381&0.381  \\
PARJ(21) &0.331&0.338 &0.366 &0.367 &0.367 \\
\hline
\end{tabular}
\newpage
\begin{tabular}{cccc}
\multicolumn{4}{c}{Table 2. The mean multiplicities of h$^-$ and participant
  proton}\\
\multicolumn{4}{c}{and the mean rapidity shift of the participant proton in}\\
\multicolumn{4}{c}{central and peripheral S+S collisions at 200A GeV/c}\\
\hline
 &N$_{h^-}$&N$_p$&$\Delta$y \\
\hline
\multicolumn{4}{l}{S+S central}\\
Data &46.8$\pm$2.5&12.8$\pm$1.4&1.58$\pm$0.15\\
LUCIAE &54.2&12.1 &1.89 \\
\multicolumn{4}{l}{S+S peripheral}\\
Data &9.8$\pm$1.0&3.1$\pm$0.8&1.0$\pm$0.15\\
LUCIAE &9.5&2.6 &1.5 \\
\hline
\end{tabular}



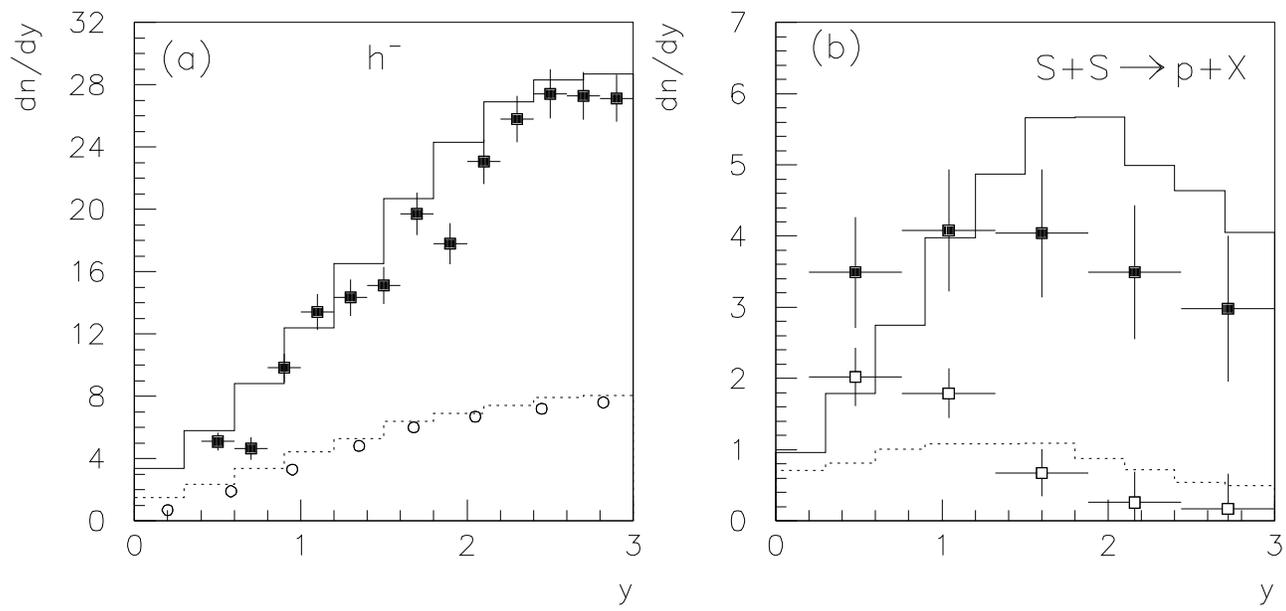
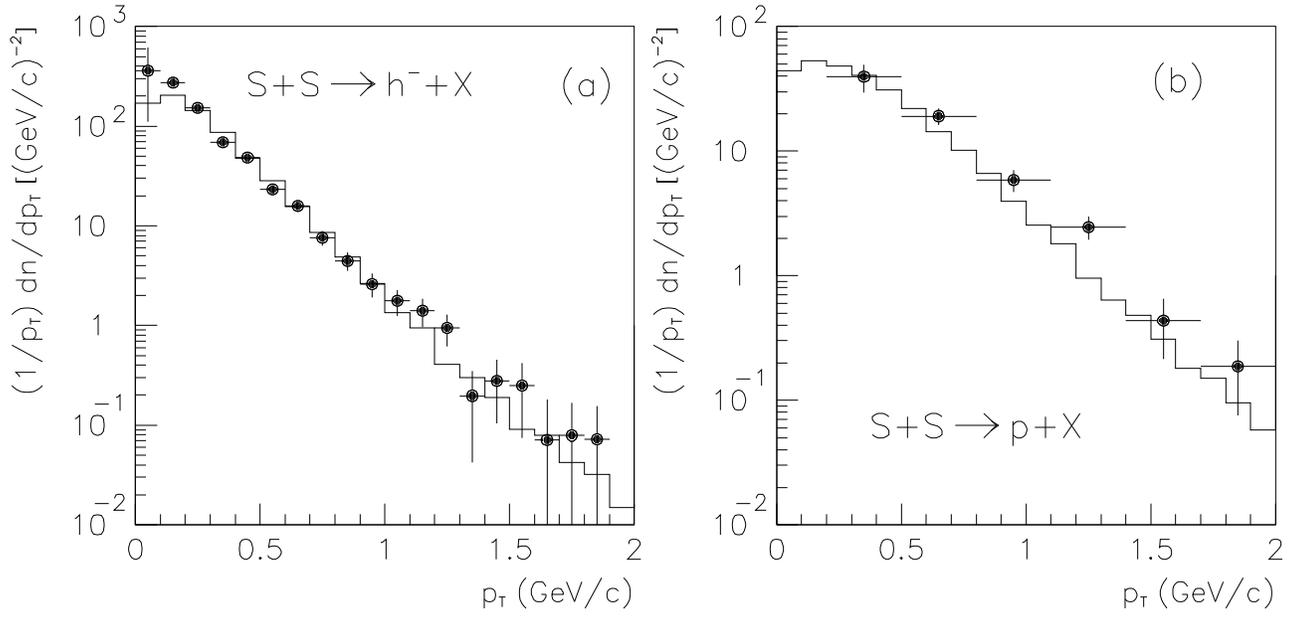
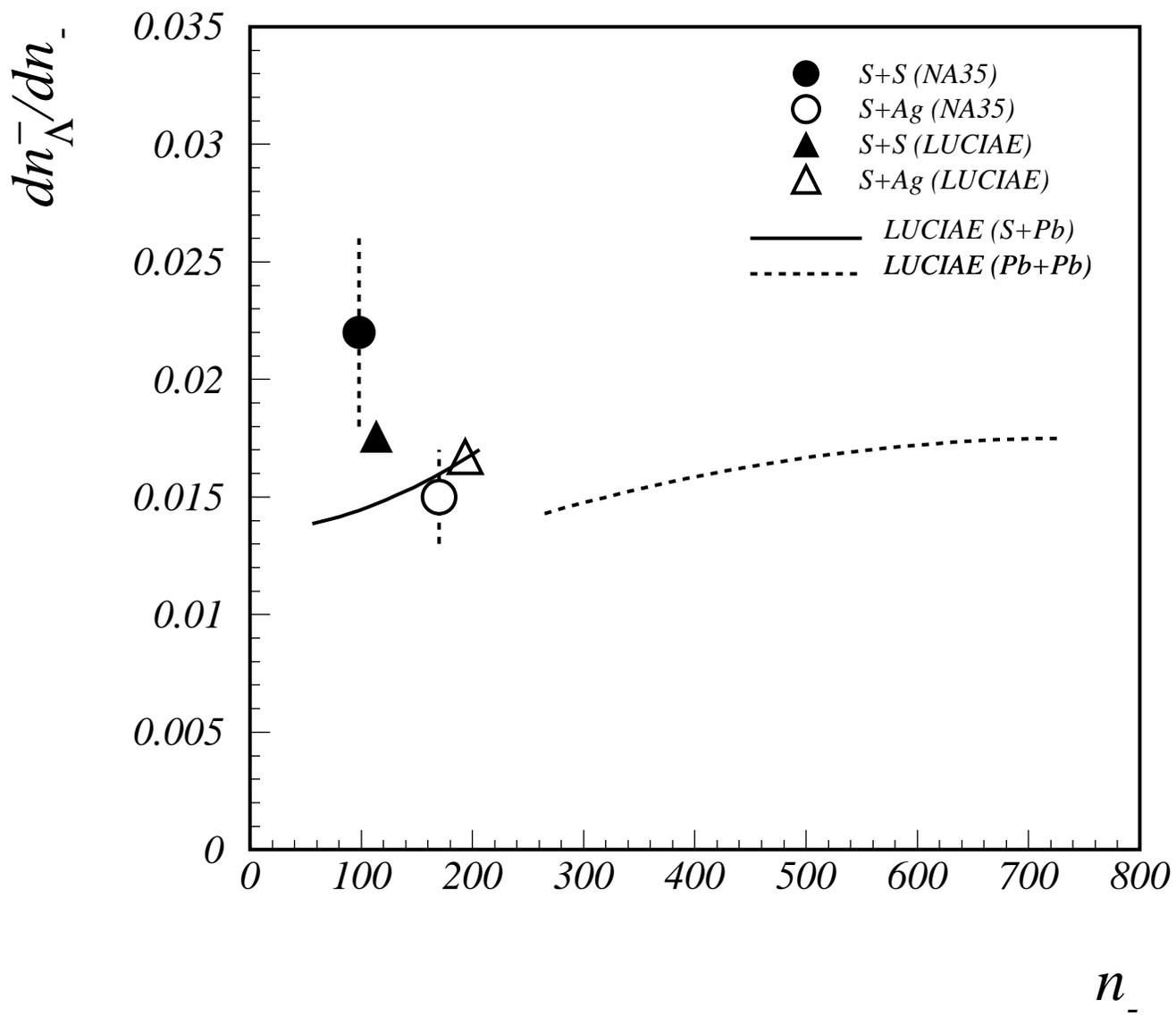
\end{document}